\documentclass[final,3p,times, onecolumn,numbers]{elsarticle}

\usepackage{natbib}
\usepackage{lineno,hyperref}
\usepackage{latexsym}
\usepackage{bm}

\usepackage{amsmath}
\usepackage{graphicx}   
\usepackage{enumerate}   
\usepackage{xcolor}

\newcommand{\be}{\begin{equation}}
\newcommand{\ee}{\end{equation}}
\newcommand{\bn}{\begin{eqnarray}}
\newcommand{\en}{\end{eqnarray}}
\newcommand{\bns}{\begin{eqnarray*}}
\newcommand{\ens}{\end{eqnarray*}}

\def\VE*{\vec{E}^{*}}

\def\ba{\begin{eqnarray}}
\def\ea{\end{eqnarray}}

\DeclareMathOperator{\sech}{sech}

\usepackage{amssymb}
\journal{Nuclear Physics B}

\begin{document}

\begin{frontmatter}



\title{Spontaneous Lorentz symmetry violation and topological defects for the Chern-Simons matter vector field}

\author{Luiz Paulo Colatto} \ead{luiz.colatto@cefet-rj.br}
\address{Centro Federal de Educa\c{c}\~ao Tecnol\'ogica Celso Suckow da Fonseca-CEFET-RJ UnED Petr\'opolis, CEP 24920-003 Rio de Janeiro, RJ, Brasil}
\author{Andr\'e Luiz Almeida Penna} \ead{andrepenna@unb.br}
%\address{Instituto de Física, Universidade de Brasília-UnB, CEP 70910-900, DF, Brasil}
\author{Wytler Cordeiro dos Santos} \ead{wytler.cordeiro@unb.br}
\address{Instituto de F\'isica, Universidade de Bras\'ilia-UnB, CEP 70910-900, DF, Brasil}

\begin{abstract}
\noindent
The study of topological defects occurring in vector and tensor fields is an intriguing subject and little explored in the literature. In this article, we analyze the topological defects arising from the spontaneous violation of Lorentz symmetry for a vector matter field with Chern-Simons term in a Minkowski spacetime in $(1+2)D$. As a consequence, the resulting nonlinear equations include a topological mass via the Chern-Simons term, which leads to a vector version of a soliton state. We show that the topological defects arising from the vector field can be categorized as either vortices, with topology $S^{1}\times\mathbb{R}$, or domain-walls, with topology $S^{0} \times\mathbb{R}^2$. The vortex solutions were analyzed using a procedure similar to the Nielsen-Olesen one,  though extended to the vector case to account for a spontaneous violation of Lorentz symmetry. We also analyze the influence of the topological mass and verify the stability of the model as well as the magnetic vortex in $(1+2)D$. We show that domain-wall solutions also emerge as an effect of the violation of the Lorentz symmetry expressed by the vacuum of the vector field. We obtain general equations for this new class of domain walls involving the Lorentz violation parameter and the topological mass. By means of the energy-momentum tensor, we verify the instability of the formation of these domain walls in $(1+2)D$.

\end{abstract}


\end{frontmatter}
%\date{\today }
%\maketitle

\section{Introduction}

According to the Standard Model, the spontaneous violation of symmetry and its consequences are well established, and gauge theories with spontaneous violation plays a central role in elementary particle theory. The existence of nontrivial vacuum expectation value directly modifies the properties of fields that couple to it and can indirectly modify them % quem é them?
through interactions with other affected fields. The vacuum value $\langle \phi \rangle$ of the Higgs field spontaneously breaks the $SU(2) \times U(1)$ gauge symmetry which is associated with the electro-weak force and generates masses for several particles, and separates the electromagnetic and weak forces. The W and Z bosons are the elementary particles that mediate the weak interaction, while the photon mediates the electromagnetic interaction. At energies much greater than 100 GeV all these particles behave in a similar manner. The Weinberg-Salam theory predicts that, at lower energies, this symmetry is broken so that the photon and the massive W and Z bosons emerge. Moreover, as we are dealing with non-linear models, it is necessary to include soliton-type configuration in order to achieve a close to actual physical states, which stability is highly dependent on the topological characteristic of the boundary conditions \cite{ryder,Weinberg}.

On the other hand, from the point of view of the vacuum at the Planck scale of the string theory, where we take a similar procedure of a spontaneous symmetry breaking (SSB) via mechanism of Higgs field, a possible spontaneous breaking of Lorentz symmetry or spontaneous Lorentz violation (LV) was detected.
The spontaneous LV a vector field, or a vector matter field,  has been extensively studied from pioneer works of V. A. Kostelecký and S. Samuel \cite{Kostelecky_Samuel} and consequently there was an increasing number of works on the theme of LV via spontaneous symmetry violation of a  vector field \cite{Kostelecky_1,Bluhm, Bertolami, Escobar}. Remarkably a vector matter field does not obey $U(1)$ gauge symmetry, or shortly, it is not a gauge field. So the vector matter field can acquires a non-zero vector vacuum expectation value inducing a spontaneous LV.

In fact studies were done on quantum gravity theories predicting the possibility of LV via the vector matter field \cite{Chkareuli}. For instance, a vector matter field $B_{\mu}$ (also called bumblebee field), coupled to gravity, with generalized quadratic kinetic terms involving up to second-order derivatives in $B_{\mu}$, and with an Einstein-Hilbert term in Riemann space-time,  is given by the Lagrangian density \cite{Bluhm2}:
\begin{eqnarray}
\label{lagrangiana_Bumblebee_1}
 {\cal L}_{B} &=& \sqrt{-g}\left[\frac{1}{2\kappa} (R - 2\Lambda) +\frac{1}{2\kappa}\xi B^{\mu}B^{\mu}R_{\mu\nu} + \rho B^{\mu}B_{\mu} R +\frac{\sigma}{2}(\nabla_{\mu}B_{\nu})(\nabla^{\mu}B^{\nu})+\frac{\tau}{2}(\nabla_{\mu}B^{\mu})(\nabla_{\nu}B^{\nu}) + \right. \nonumber \\  && \left. -\frac{1}{4} B_{\mu\nu}B^{\mu\nu} - V(B_{\mu}B^{\mu}) \right] \, ,
\end{eqnarray}
where $R$ is the curvature scalar, $R_{\mu\nu}$ is the Ricci tensor, $\kappa = 16\pi G$ and $\xi$,  $\rho$, $\sigma$ and $\tau$ are parameters.
The  vector matter field strength or kinetic term is defined as
\begin{equation}
 B_{\mu\nu} =\nabla_{\mu}B_{\nu} - \nabla_{\nu}B_{\mu} = \partial_{\mu}B_{\nu} - \partial_{\nu}B_{\mu}.
\end{equation}
The potential $V$ providing a non-zero vacuum expectation value for vector field  $B_{\mu}$ has the following function form
\begin{equation}
\label{potential}
V(B_{\mu}B^{\mu})=  \frac{\lambda}{4}\left(B_{\mu}B^{\mu} \pm b^2 \right)^2.
\end{equation}
Then, the non-zero vacuum expectation value of the vector matter field is obtained by
\begin{equation}
\label{minimal_potential}
 \frac{\partial V}{\partial B_{\mu}} =  \frac{\lambda}{2}\left(B_{\nu}B^{\nu} \pm b^2 \right)B^{\mu} = 0
\end{equation}
Therefore it is solved when the field $B^{\mu}$ acquires a non-zero vacuum expectation value,
\begin{equation}
\langle 0|B^{\mu}|0 \rangle = b^{\mu},
\end{equation}
where $b^{\mu}$ has constant magnitude but different orientation at different space-time points, and therefore the global Lorentz and translation symmetries are broken \cite{Bluhm}. It is remarkable that the potential  has a minimum with respect to its argument or is constrained to zero when
\begin{equation}
B^{\mu}B_{\mu} = \pm b^2.
\end{equation}
So, we state that the magnitude of the vector field is a time-like vector when we have $+b^2$ and the vector field is a space-like vector when we have $-b^2$. The LV yields by a vector field can result in topological defects as pointed out by Seifert in Ref. \cite{Seifert}.

%%%%%%%%%%%%%%%%%%%%%%%%%%%%%%%%%%%%%%%%%%%%%%%
%%%%%%%%%%%%%%%%%%%%%%%%%%%%%%%%%%%%%%%%%%%%%%%
%%%%%%%%%%%%%%%%%%%%%%%%%%%%%%%%%%%%%%%%%%%%%%%
%%%%%%%%%%%%%%%%%%%%%%%%%%%%%%%%%%%%%%%%%%%%%%%
%%%%%%%%%%%%%%%%%%%%%%%%%%%%%%%%%%%%%%%%%%%%%%%
%%%%%%%%%%%%%%%%%%%%%%%%%%%%%%%%%%%%%%%%%%%%%%%
%%%%%%%%%%%%%%%%%%%%%%%%%%%%%%%%%%%%%%%%%%%%%%%
%%%%%%%%%%%%%%%%%%%%%%%%%%%%%%%%%%%%%%%%%%%%%%%
On the other hand, LV in $(2+1)$-dimensional field theories, often is induced by a fixed background vector $v^\mu$, which furnishes an interesting phenomenological structure with important implications for condensed matter systems. This violation can be introduced via a Chern-Simons-like coupling term $\varphi \epsilon_{\mu\nu\kappa} v^\mu \partial^\nu A^\kappa$ that could be derived from dimensional reduction of the Carroll-Field-Jackiw model \cite{Belich2003}, which modifies the propagators of the gauge and scalar fields, introducing new poles and raising causality concerns for timelike $v^\mu$, though unitarity is generally preserved for spacelike backgrounds. The low-energy electron-electron interaction potential in such models becomes attractive due to scalar and gauge contributions \cite{Ferreira2004}, a crucial feature for theorizing Cooper pair formation in planar systems. Furthermore, LV terms can be generated radiatively through fermion loops with non-minimal couplings \cite{Bufalo2014} or non-perturbatively via the condensation of topological defects using the Julia-Toulouse approach \cite{Nascimento2014}. These mechanisms are deeply connected to the induction of mixed and conventional Chern-Simons terms, which are central to describing $T$-invariant charge fractionalization, as in graphene, or $T$-violation scenarios~\cite{Charneski2009}. Notably, at finite temperature, LV coupling induces a non-trivial magnetic screening mass \cite{Bufalo2014}, modifying the infrared behavior and ensuring a finite fermion self-energy. Thus, LV in three dimensions provides a powerful effective framework for exploring anisotropic phenomena, fractional statistics, and potential superconductivity in quantum materials.

%%%%%%%%%%%%%%%%%%%%%%%%%%%%%%%%%%%%%%%%%%%%%%%
%%%%%%%%%%%%%%%%%%%%%%%%%%%%%%%%%%%%%%%%%%%%%%%
%%%%%%%%%%%%%%%%%%%%%%%%%%%%%%%%%%%%%%%%%%%%%%%
%%%%%%%%%%%%%%%%%%%%%%%%%%%%%%%%%%%%%%%%%%%%%%%
%%%%%%%%%%%%%%%%%%%%%%%%%%%%%%%%%%%%%%%%%%%%%%%
%%%%%%%%%%%%%%%%%%%%%%%%%%%%%%%%%%%%%%%%%%%%%%%
%%%%%%%%%%%%%%%%%%%%%%%%%%%%%%%%%%%%%%%%%%%%%%%
%%%%%%%%%%%%%%%%%%%%%%%%%%%%%%%%%%%%%%%%%%%%%%%

Remarkably many physical effects, in particular topological defects, are conjectured to have a LV origin, and  their theoretical developments have been largely worked \cite{Chkareuli,and1,and2,bar,Bazeia,MFerreira,Passos}.  In fact, in $(1+2)D$ models framework we can observe the presence of topological effects which can be theoretically represented through the vacuum of a $(1+2)D$ model with Chern-Simons photon coupling \cite{jackiw2,jackiw3}.

%%%%%%%%%%%%%%%%%%%%%%%%%%%%%%%%%%%%%%%%%%%%%%%
%%%%%%%%%%%%%%%%%%%%%%%%%%%%%%%%%%%%%%%%%%%%%%%
%%%%%%%%%%%%%%%%%%%%%%%%%%%%%%%%%%%%%%%%%%%%%%%
%%%%%%%%%%%%%%%%%%%%%%%%%%%%%%%%%%%%%%%%%%%%%%%
%%%%%%%%%%%%%%%%%%%%%%%%%%%%%%%%%%%%%%%%%%%%%%%
%%%%%%%%%%%%%%%%%%%%%%%%%%%%%%%%%%%%%%%%%%%%%%%
%%%%%%%%%%%%%%%%%%%%%%%%%%%%%%%%%%%%%%%%%%%%%%%
%%%%%%%%%%%%%%%%%%%%%%%%%%%%%%%%%%%%%%%%%%%%%%%
Our purpose in this work is to analyse two topological defects that arise from LV originated from non-zero vector vacuum expectation values, namely the domain wall and a stable $(1+2)D$ vortex lines solutions, starting from interacting vector matter field with the $U(1)$ gauge vector field.  We will discuss the contribution of the Chern-Simons-type terms to asymptotic behaviour of the vector field.

The outline of the work is the following:
In Sec. II we treat the classification of the vacuum manifold defects due to a vector in order to originated LV.
Sec. III is devoted to analyse the topological defect domain wall in $(1+2)D$ spacetime.
Sec. IV we study the second topological defect, the vortex lines.
Finally, in the Conclusion we discuss the results and give perspectives of new approaches to the subject.

%%%%%%%%%%%%%%%%%%%%%%%%%%%%%%%%%%%%%%%%%
%%%%%%%%%%%%%%%%%%%%%%%%%%%%%%%%%%%%%%%%%
%%%%%%%%%%%%%%%%%%%%%%%%%%%%%%%%%%%%%%%%%
%%%%%%%%%%%%%%%%%%%%%%%%%%%%%%%%%%%%%%%%%
%%%%%%%%%%%%%%%%%%%%%%%%%%%%%%%%%%%%%%%%%
%%%%%%%%%%%%%%%%%%%%%%%%%%%%%%%%%%%%%%%%%

\section{The matter vector field formulation and classification of the vacuum manifold defects}

It is worth to remark that vector matter models are not gauge invariant \cite{Bluhm2, Escobar}, so we assume that the vector matter field plays the role of matter field in Minkowski space-time. Furthermore we add the condition that vector matter field is electrically charged, i.e., we assume a charged complex vector field that maintain the global $U(1)$ invariance.
For our proposals we are going to take the parameters $\xi$, $\rho$ and $\sigma$ of the equation \eqref{lagrangiana_Bumblebee_1} to be null, as well as $g_{\mu\nu}=\eta_{\mu\nu} = \mbox{diag}(+--)$ with $\sqrt{g} = 1$, so that the Lagrangian  (\ref{lagrangiana_Bumblebee_1}) is modified as follows,
\begin{equation}
\label{lagrangiana_Bumblebee_2}
 {\cal L}_{B} = -\frac{1}{2} B_{\mu\nu}^{*}B^{\mu\nu} -(\partial_{\mu}B^{\mu})^{*}(\partial_{\nu}B^{\nu}) - \frac{\lambda}{4}\left(B_{\mu}B^{\mu} \pm b^2 \right)^2.
\end{equation}
We use the charged matter vector field model and we assume it is living in a $(1+2)D$ world, including global $U(1)$ invariant terms, and topological Chern-Simons-type terms. It can be written as,
\begin{equation}
\label{Lag1}
{\cal L}_{BCS}=-\frac{1}{2}B_{\mu\nu}^{*}B^{\mu\nu} -
(\partial_{\mu}B^{\mu})^{*}(\partial_{\nu}B^{\nu})+m\epsilon^{\mu\nu\kappa}B_{\mu}\partial_{\nu}B_{\kappa}^*+m\epsilon^{\mu\nu\kappa}B_{\mu}^*\partial_{\nu}B_{\kappa}
 -\frac{\lambda}{4}\left(B_{\mu}B^{\mu} \pm b^2 \right)^2 ,
\end{equation}
where $B_{\mu}$ is a charged vector field and $B_{\mu\nu}=\partial_{\mu}B_{\nu}-\partial_{\nu}B_{\mu}$ is the vector matter field strength. The global symmetry $U(1)$ allows us to insert a topological massive Chern-Simons-type term that does not play a role in the vacuum achievement but will be important to the mass term definition in the asymptotic limit behaviour that will be made clear further ahead.
We can verify that this model gives rise to a spontaneous symmetry violation mechanism whose non-trivial vacuum
%\cite{and1,and2,matter}
allows two particular solutions from the analyses of topological defects originated of LV of the Lagrangian (\ref{Lag1}).

%%%%%%%%%%%%%%%%%%%%%%%%%%%%%%%%%%%%%%%%%%%%%%
%%%%%%%%%%%%%%%%%%%%%%%%%%%%%%%%%%%%%%%%%%%%%%

On the other hand, we must remark that in the framework of $D=1+3$, T. W. B. Kibble in reference \cite{Kibble} have demonstrated that the topology of the vacuum manifold $\cal M$ determines whether defects appear at a particular symmetry violation. Domain wall, in this framework, can be formed if $\cal M$ has disconnected components, strings can be formed if $\cal M$ is not simply connected, that is, if it contains unshrinkable loops, and monopoles  can be formed if $\cal M$ contains  unshrinkable surfaces. The relevant properties of a generic manifold, $\cal M$, are most conveniently studied using homotopy theory: the $n^{\mbox{\small th}}$ homotopy group $\pi_{n}({\cal M})$ classifies qualitatively distinct mappings from the $n$-dimensional sphere $S^{n}$ into the manifold $\cal M$.

Let us first consider a theoretical model in which the vacuum manifold $\cal M$ is not simply connected, that is a manifold that contain at least a hole in it and some loops can not be continuously shrunk to a point. This topological property have been established through the first homotopy group $\pi_{1}({\cal M})$. If the manifold $\cal M$ is not simply connected, $\pi_{1}({\cal M})$ is not trivial and the topological defect, the string structure  will occur. The Nielsen-Olesen vortex-line is an example of a solution of string, in which case the manifold is ${\cal M} = S^{1} \times \mathbb{R}^{2}$ with $\pi_{1}(S^{1} \times \mathbb{R}^{2})=\pi_{1}(S^{1}) = \mathbb{Z}$ non-trivial.
If the vacuum manifold $\cal M$  has a non-trivial second homotopy group $\pi_{2}({\cal M})$, that is, two-surfaces can not be continuously shrunk to a point, i.e., the manifold of degenerate vacua contains non-contractible two-surfaces, like the sphere $S^{2}$, then a topological defect arises, that is a monopole solution. The simplest monopole solution is the 't Hooft-Polyakov monopole.
Another topological defect arises with the breaking of a discrete symmetry, when the vacuum manifold $\cal M$ consists of several disconnected components, domain walls will occur. The notation used for this topological defect is $\pi_{0}({\cal M})$, and this is not strictly speaking a homotopy group, it has no group structure but represents merely the number of connected components of manifold
$\cal M$ in accordance with reference \cite{Hilton} on page 276.
A thorough description of topological defects of  the vacuum manifold in  physics can be found in reference \cite{Vilenkin}.

In this work we purpose the spontaneous LV of the vacuum via the matter vector field and we following the methods indicated by M.D. Seifert in reference \cite{Seifert} which is based on a bumblebee model in $D=1+3$ space-time dimension, to develop a similar analysis to $D=1+2$ framework. Thus we can display the vector norm in orthonormal Minkowski frame  with the signature  $(+,-,-)$ as,
\begin{equation}
 \label{vector_norm_1}
 b_{\mu}b^{\mu} = (b_{0})^2 - (b_{1})^2 - (b_{2})^2 = \pm |b|^2.
\end{equation}
One must observe both possibilities, the time-like vector or the space-like vector in a spontaneous LV. We can determine the topological defect of the vacuum manifold for time-like vector whose components satisfy
\begin{equation}
 \label{vector_norm_2}
 (b_{0})^2 = (b_{1})^2 + (b_{2})^2 + |b|^2.
\end{equation}
This space of time-like vectors has topology $S^{0}\times \mathbb{R}^{2}$ thus  $\pi_{0}(S^{0}\times \mathbb{R}^{2}) =\pi_{0}(S^{0})$. This spontaneous LV gives rise to a domain wall as a topological defect. On the other hand, the   component space of the space-like vectors satisfy
\begin{equation}
 \label{vector_norm_3}
 (b_{1})^2 + (b_{2})^2 = (b_{0})^2 + |b|^2.
\end{equation}
Then, the space of space-like vectors has topology $S^{1}\times \mathbb{R}$ and the first homotopy group of this vacuum manifold is $\pi_{1}(S^{1})= \mathbb{Z} $, which is non-trivial. This spontaneous LV gives rise to vortex lines as a topological defect, very similar to the Nielsen-Olesen type approach \cite{Nielsen}.

\section{Domain Wall}

The domain wall is a topological defect that comes from the breaking of vacuum symmetry of the potential (\ref{potential}) when the matter vector field is time-like, and consequently the vacuum manifold consists of disconnected components \cite{Vilenkin, Seifert}. The equation of motion that comes from the Lagrangian (\ref{Lag1}) is,
\begin{equation}
\label{motion_equation_B_01}
\partial_{\nu}\partial^{\nu} B^{\mu} + 2 m\epsilon^{\mu\nu\kappa}\partial_{\nu}B_{\kappa} =  \frac{\lambda}{2}\left(B_{\nu}^{*}B^{\nu} - b^2 \right) B^{\mu}.
\end{equation}
So, we obtain a time-like matter vector field solution that satisfies $B_{\mu}^{*}B^{\mu} =  b^{2}$ when vacuum symmetry is broken. In accordance with the basic description of domain walls, we assume that $B^{\mu}$ is an stationary vector field that depends only on the variable $x$, i.e.,  $B^{\mu} = B^{\mu}(x)$. Therefore from the equation (\ref{motion_equation_B_01}) we can obtain three equations of motion displayed as follows,
\begin{eqnarray}
\label{motion_equation_B_02}
 \frac{d^2 B^{t}}{dx^2} - 2 m\frac{d B^{y}}{dx} &=& -\frac{\lambda}{2}\left(B_{\nu}^{*}B^{\nu} - b^2 \right)  B^{t},
\label{motion_equation_B_03} \\
 \frac{d^2 B^{x}}{dx^2}  &=& -\frac{\lambda}{2}\left(B_{\nu}^{*}B^{\nu} - b^2 \right)  B^{x},
\label{motion_equation_B_04} \\
 \frac{d^2 B^{y}}{dx^2} + 2 m\frac{d B^{t}}{dx} &=& -\frac{\lambda}{2}\left(B_{\nu}^{*}B^{\nu} - b^2 \right)  B^{y}.
\end{eqnarray}
In order to obtain a time-like vector field solution we also assume that
\begin{equation}
\label{field-domain-wall_01}
 B^{\mu}(x) = (B^{t},B^{x},B^{y}) = (f(x), 0 , -K),
\end{equation}
where $K$ is a constant. The system of three equations reduces to a system of two equations then,
\begin{equation}
\label{motion_equation_f_01}
 \frac{d^2 f(x)}{dx^2} = - \frac{\lambda}{2}\left(B_{\nu}^{*}B^{\nu} - b^2 \right) f(x),
\end{equation}
and
\begin{equation}
\label{motion_equation_f_02}
2 m\frac{d f(x)}{dx} = - \frac{\lambda}{2}\left(B_{\nu}^{*}B^{\nu} - b^2 \right) (-K).
\end{equation}
The term $-\frac{\lambda}{2}\left(B_{\nu}^{*}B^{\nu} + b^2 \right)$ can be isolated in equation (\ref{motion_equation_f_02}) to use it in the equation (\ref{motion_equation_f_01}), we obtain the following  differential equation,
\begin{equation}
 \frac{d^2 f(x)}{dx^2} + \frac{2m}{K} f(x) \frac{d f(x)}{dx} = 0,
\end{equation}
whose solution is $f(x) = \sqrt{\frac{KC}{m}} \tanh\left(\sqrt{\frac{mC}{K}}~x\right)$ where $C$ is a integration constant which can be determined when we directly apply the function $f(x)$ to the equations (\ref{motion_equation_f_01}) and (\ref{motion_equation_f_02}).
Therefore we obtain that $C=\frac{2m^2}{\sqrt{-\lambda}} + \frac{b^2\sqrt{-\lambda}}{2}$ and consequently $K = \frac{2m}{\sqrt{-\lambda}}$.  Resulting in
\begin{equation}
\label{solution_domain_wall_01}
 f(x) = \sqrt{b^2 - \frac{4m^2}{\lambda}} ~ \tanh\left(\sqrt{m^2 - \frac{\lambda b^2}{4}}~x\right),
\end{equation}
It is easy to check if we take the limit as $x\rightarrow \pm \infty$, we have $B_{\mu}^{*}B^{\mu}= b^2$.
The expression $(\ref{solution_domain_wall_01})$ is a well-known solution of domain wall \cite{Vilenkin,Seifert}.

The next point to check about this solution concerns the contribution to the energy-momentum tensor that can be found directly from the variation of action of  the Lagrangian (\ref{Lag1}) with respect to metric tensor in a three-dimensional manifold $\cal M$. We can remark that the Chern-Simons topological  action is independent of the spacetime metric, and therefore the Lagrangian density of Chern-Simons does not contribute to the energy-momentum tensor \cite{Dunne}. Thus, in accordance with reference \cite{Landau} and taking the signature $(+--)$, we can calculate the energy-momentum tensor with the action of the Lagrangian (\ref{lagrangiana_Bumblebee_2}) in such way that
\begin{equation}
\label{energy-momentum-tensor_01}
 \delta \left(\int dt~d^2x~\sqrt{g}~{\cal L}_{B} \right) = \frac{1}{2} \int  dt~d^2x~\sqrt{g}~T_{\mu\nu}\delta g^{\mu\nu}.
\end{equation}
and then bearing in mind that we are dealing with a $(1+2)D$ Minkowski space, so we obtain the energy momentum tensor as:
\begin{equation}
\label{energy-momentum-tensor_02}
 T_{\mu\nu} = - B_{\mu\rho}^{*}{B_{\nu}}^{\rho} - \frac{\lambda}{2}\left(B_{\nu}^{*}B^{\nu} + b^2 \right) B_{\mu}^{*}B_{\nu} - \left(\partial_{\mu}B_{\nu}^{*}\right)\left(\partial_{\kappa}B^{\kappa}\right) - \left(\partial_{\mu}B_{\nu}\right)\left(\partial_{\kappa}B^{\kappa}\right)^{*} + c.c. - \eta_{\mu\nu} {\cal L}_{B},
\end{equation}
where $c.c.$ means complex conjugate.
It is necessary to carefully observe that the Lagrangian density (\ref{lagrangiana_Bumblebee_2}) lives in a $(1+2)D$ Minkowski space, and it has dimension of surface energy density (energy divided by unit area).
Therefore we use the solution (\ref{solution_domain_wall_01}) in the equation (\ref{field-domain-wall_01}) to yield the components of the energy-momentum tensor
\begin{eqnarray}
 T_{00} &=& - 4\left(\frac{2m^2}{\sqrt{-\lambda}}+\frac{b^2\sqrt{-\lambda}}{2} \right)^2\tanh^2(u)\sech^2(u),\cr
 T_{22} &=& 2\left(\frac{2m^2}{\sqrt{-\lambda}}+\frac{b^2\sqrt{-\lambda}}{2} \right)^2\sech^4(u) + \left(16\frac{m^4}{\lambda} - 4b^2m^2 \right)\sech^2(u), \cr
 T_{02} &=&  \frac{2 m}{\sqrt{-\lambda}} \left(b^2\lambda - 4m^2 \right)^{3/2}\sech^2(u)\tanh(u), \cr
 T_{11} &=& T_{12} = T_{01} = 0.
\end{eqnarray}
where $u=\sqrt{m^2 - \frac{b^2 \lambda}{4}} ~ x$.
We can observe that if we take the factor $m$ as zero, all above components of the energy-momentum tensor are reduced to the results achieved by Seifert in the reference \cite{Seifert}. Nonetheless, we remark that the above components of the energy-momentum tensor live in a $(1+2)D$ spacetime, so the component $T_{00}$ is already a surface energy density. The wall tension in the tangential direction $y$ is the component $T_{22}$.

We observe that the density energy $T_{00}$ is negative and it does not satisfy the energy condition. In this situation the energy stability for a domain wall in $(1+2)D$ spacetime become a puzzle. Furthermore, a domain wall in $(1+2)D$ spacetime is not a wall with tangential tension on the two directions on the surface \cite{Vilenkin}.
In this way, it seems that domain wall is not possible to obtain as a legitimate defect on this scenario.

%%%%%%%%%%%%%%%%%%%%%%%%%%%%%%%%%%%%
%%%%%%%%%%%%%%%%%%%%%%%%%%%%%%%%%%%%
%%%%%%%%%%%%%%%%%%%%%%%%%%%%%%%%%%%%
%%%%%%%%%%%%%%%%%%%%%%%%%%%%%%%%%%%%
%%%%%%%%%%%%%%%%%%%%%%%%%%%%%%%%%%%%
%%%%%%%%%%%%%%%%%%%%%%%%%%%%%%%%%%%%

\section{Vortex lines}

%%%%%%%%%%%%%%%%%%%%%%%%%%%%%%%%
%%%%%%%%%%%%%%%%%%%%%%%%%%%%%%%%
%%%%%%%%%%%%%%%%%%%%%%%%%%%%%%%%
%%%%%%%%%%%%%%%%%%%%%%%%%%%%%%%%
%%%%%%%%%%%%%%%%%%%%%%%%%%%%%%%%
%%%%%%%%%%%%%%%%%%%%%%%%%%%%%%%%
%%%%%%%%%%%%%%%%%%%%%%%%%%%%%%%%
%%%%%%%%%%%%%%%%%%%%%%%%%%%%%%%%

%%%%%%%%%%%%%%%%%%%%%%%%%%%%%%%%
%%%%%%%%%%%%%%%%%%%%%%%%%%%%%%%%
%%%%%%%%%%%%%%%%%%%%%%%%%%%%%%%%
%%%%%%%%%%%%%%%%%%%%%%%%%%%%%%%%
%%%%%%%%%%%%%%%%%%%%%%%%%%%%%%%%
%%%%%%%%%%%%%%%%%%%%%%%%%%%%%%%%
%%%%%%%%%%%%%%%%%%%%%%%%%%%%%%%%
%%%%%%%%%%%%%%%%%%%%%%%%%%%%%%%%
Studies on vortex solutions within the LV framework of the Standard Model Extension have revealed a rich spectrum of topological defects with novel properties. We can briefly comment on some vortex solutions such as those that occur in the CPT-even sector. LV leads to uncharged Bogomol'nyi, Prasad, Sommerfeld (BPS) vortices that can exhibit compact like profiles, where the LV parameter controls the size of the defect's core, and can also induce fractional quantization of the magnetic flux when LV terms are present in the Higgs sector \cite{Miller2012}. Furthermore, the CPT-odd sector allows for the existence of electrically charged BPS vortices even in the absence of the Chern-Simons term, with the LV background coupling the electric and magnetic sectors; these charged solutions show localized magnetic flux and electric field reversal, a phenomenon of potential interest for condensed matter systems like multi-component superconductors \cite{Casana2012}. This behaviour is also prominently featured in a dimensionally reduced Maxwell-Chern-Simons model, where a CPT-odd LV parameter can cause a pronounced negative magnetic flux near the origin for sufficiently high winding numbers, while the total flux remains quantized~\cite{Casana2014}. These studies indicate LV as an important mechanism for generating vortices with controllable features such as size, charge, and anomalous flux that are unattainable in standard Lorentz invariant theories, all while preserving the fundamental BPS property of energy proportional to the quantized magnetic flux.

%%%%%%%%%%%%%%%%%%%%%%%%%%%%%%%%
%%%%%%%%%%%%%%%%%%%%%%%%%%%%%%%%
%%%%%%%%%%%%%%%%%%%%%%%%%%%%%%%%
%%%%%%%%%%%%%%%%%%%%%%%%%%%%%%%%
%%%%%%%%%%%%%%%%%%%%%%%%%%%%%%%%
%%%%%%%%%%%%%%%%%%%%%%%%%%%%%%%%
%%%%%%%%%%%%%%%%%%%%%%%%%%%%%%%%
%%%%%%%%%%%%%%%%%%%%%%%%%%%%%%%%

As we have already mentioned previously, the vortex lines structures can be derived from the non-trivial topology of the vacuum of the space-like vector field, when it  has topology $S^{1}\times \mathbb{R}$ in $(1+2)D$. The spontaneous LV gives rise to vortex lines topological defect, very similar to the Nielsen-Olesen vortex lines \cite{Nielsen}.
The potential of the space-like vector field in equation (\ref{potential}) must be
\begin{equation}
\label{potential_space-like}
 V(B_{\mu}^{*}B^{\mu})=\frac{\lambda}{4}\left(B_{\mu}^*B^{\mu} +b^2\right)^2=0,
\end{equation}
where $B_{\mu}^*B^{\mu} = -b^{2}$, when spontaneous LV happens. In the same way of the Nielsen-Olesen approach, let us first look at the issue of the energy under static conditions.
From the non-trivial topology of the vacuum in $(1+2)D$ and in the same format as the abelian-Higgs model \cite{ryder,Vilenkin}, we consider the time-component and the space-component of the matter vector  written as,
\begin{equation}
\label{Bmu}
B_0=b_0e^{in\theta},\hspace{2cm}
B_r=b_r e^{in\theta}, \hspace{2cm}  B_{\theta}=b_{\theta} e^{in\theta} \, ,
\end{equation}
where $b_0$ is a time component, $b_r$ and $b_{\theta}$ are radial and angular vector components of the LV parameter $b_{\mu}$.
In our case, the $B_{\mu}$ field generates a vortex whose stability is analysed starting from static configuration for the Hamiltonian functional density,
$\dot{B}_{\mu} = 0$. In static configuration we have that the Hamiltonian reduces to ${\cal H}_{s.c.}=-{\cal L}_{s.c.} = \frac{1}{2} B_{\mu\nu}^{*}B^{\mu\nu} + (\partial_{\mu}B^{\mu})^{*}(\partial_{\nu}B^{\nu})$, so that,
\begin{equation}
\label{Hamilt1}
{\cal H}_{s.c.}=-(\vec \nabla B_0)^{*}\cdot(\vec \nabla B_0)+(\vec \nabla
\times \vec B)^*\cdot(\vec \nabla
\times \vec B)+(\vec \nabla
\cdot \vec B)^*\cdot(\vec \nabla
\cdot \vec B),
\end{equation}
where we have also admitted a static configuration for the spontaneous LV in the potential (\ref{potential_space-like}), where $V(B_{\mu},B_{\mu}^*)=0$.
One more time, the topological Chern-Simons-type term in (\ref{Lag1}) does not contribute to the Hamiltonian density \cite{Dunne}.
Let us now treat the Hamiltonian (\ref{Hamilt1}) in three dimensions, admitting a cylindrical symmetry around the $z$-axis. Thus, the $z$-component of $B_\mu$ is assumed to be constant  and exterior to space $S^{1}\times \mathbb{R}$ in $(1+2)D$, and from (\ref{Bmu}) we have found that,
\begin{eqnarray}
\label{df}
\vec \nabla B_0=
\frac{inb_0}{r}\,e^{in\theta}\hat\theta,\hspace{1cm}
\vec \nabla \cdot \vec B= \frac{1}{r}(b_r+inb_{\theta})e^{in\theta},
\hspace{1cm}
\vec \nabla \times \vec B= \frac{1}{r}(b_{\theta}-inb_r)e^{in\theta}\hat z.
\end{eqnarray}
Hence substituting (\ref{df}) with (\ref{Hamilt1}) we find that at $r \rightarrow \infty$ the Hamiltonian density is given by,
\begin{equation}
\label{Hamilt2}
{\cal H}_{s.c.}=\frac{-n^2b_0^2+(n^2+ 1)(b_r^2+b_{\theta}^2)}{r^2}
\end{equation}
and the energy of the matter vector field relative to the topological vacuum defect is,
\begin{eqnarray}
\label{Energia}
E=\int^{\infty}{\cal
H}_{s.c.}rdrd\theta=\left[-n^2b_0^2+(n^2+1)\left(b_r^2+b_{\theta}^2\right)\right]
\ln|r|\Bigr|^{\infty}.
\end{eqnarray}
As the abelian-Higgs model of Nielsen-Olesen approach, we treat this logarithmic divergence by adding a gauge field $A_{\mu}$ in the model, so,
\begin{equation}
\label{DC}
D_{\mu}B_{\nu}=\partial_{\mu}B_{\nu}+ieA_{\mu}B_{\nu} \, .
\end{equation}
We shall then assume the gauge choice $A_{0}=0$ and $\vec{A}= A^{r}\hat r + A^{\theta}\hat{\theta} = \frac{1}{e}\vec \nabla(n\theta)$, for very large $r$ value. As is well known, we write it analytically in cylindrical symmetry, or
\begin{equation}
\label{PG}
A_r \rightarrow 0 \hspace{2cm}\mbox{and}\hspace{2cm} - A^{\theta} =  A_{\theta} ,
\rightarrow -\frac{n}{er}
\end{equation}
since we treat the covariant derivative (\ref{DC}) at infinity, the results is
\begin{equation}
D_{\mu}B_{\nu}=0.
\end{equation}
Thus, the stability of this model requires a ``vector electrodynamics'' whose Lagrangian is defined by,
\begin{equation}
\label{Lag2}
{\cal L}=-\frac{1}{4}f_{\mu\nu}f^{\mu\nu}-\frac{1}{2}{\cal B}_{\mu\nu}^{*}{\cal B}^{\mu\nu} -(D_{\mu}B^{\mu})^{*}(D_{\nu}B^{\nu})+m\epsilon^{\mu\nu\kappa}B_{\mu}^*(D_{\nu}B_{\kappa})+m\epsilon^{\mu\nu\kappa}B_{\mu}(D_{\nu}B_{\kappa})^*
 - V(B_{\mu}^{*}B^{\mu}),
\end{equation}
where ${\cal B}_{\mu\nu}= D_{\mu}B_{\nu}-D_{\nu}B_{\mu}$ and $f_{\mu\nu}=\partial_{\mu}A_{\nu}-\partial_{\nu}A_{\mu}$. From the relations (\ref{PG}) which define a pure gauge, the Lagrangian model (\ref{Lag2}) assumes a finite energy configuration for its Hamiltonian functional density at infinity,
\begin{equation}
{\cal B}_{\mu\nu}\rightarrow 0 \hspace{2cm}\mbox{and}\hspace{2cm} {\cal H}\rightarrow 0
\end{equation}
and the gauge field $A_{\mu}$ gives rise to a soliton magnetic flux or vortex,
\begin{equation}
\Phi=\oint \vec A \cdot d\vec l= \frac{2\pi n}{e},
\end{equation}
which is analogous to the  abelian-Higgs model of Nielsen-Olesen  and represents a magnetic flux quantization as in the London and Ginzburg-Landau equations to describe the type-II superconductors \cite{ryder}.

Based on the Lagrangian \eqref{Lag2}, we now analyse the emergence of soliton kind solutions based on the Nielsen-Olesen work \cite{Nielsen}, namely vortices, with topological mass contribution of the vector matter field. In fact, it yields a vortex solution whose dynamical equation for $A_{\mu}$ is written as
\bn
\label{em}
ie(B_{\nu}^*{B}^{\mu\nu}-B_{\nu}{B}^{\mu\nu *})+ ie
(B^{\mu*}\partial_{\nu}B^{\nu}-B^{\mu
}\partial_{\nu}B^{\nu*})-iem\epsilon^{\mu\nu\kappa}(B_{\nu}^*B_{\kappa}-B_{\nu}B_{\kappa}^*)-2e^2A^{\mu}(B_{\nu}^*B^{\nu})=\partial_{\nu}f^{\mu\nu},
\en
and the dynamical equation for $B_{\mu}$ is given by
\begin{equation}
\label{em0} D_{\mu}D^{\mu}B^{\nu}+2m\epsilon^{\nu\kappa\lambda}D_{\kappa}B_{\lambda}= \frac{\lambda}{2}\left(B_{\mu}^*B^{\mu}+b^2\right)B^{\nu} \, .
\end{equation}
We are seeking for vortex solutions in the system described by (\ref{em}). In order to extract further information about the system we assume cylindrical coordinates in such way that
\be
\label{AeB}
A^{\mu} = (0,0,A^{\theta})=(0,0,A(r)) \hspace{2cm} \mbox{and} \hspace{2cm} B_{\mu}=\beta_{\mu}(r)e^{in\theta},
\ee
where the asymptotic behaviours of $\beta_{\mu}(r)$ are given by,
\bn
\label{limit}
\lim_{r\rightarrow\infty}\beta_{\mu}(r)= b_{\mu} \hspace{3cm}\lim_{r\rightarrow 0}\beta(r)= 0 \, ,
\en
with $\mu=0,\,r,\,\theta$  which are, respectively, the time, $r$ and $\theta$ cylindrical components of the vector matter field. So $B_{\mu}^*B^{\mu}=\beta_{\mu}(r)\beta^{\mu}(r) = \beta(r)^2$ in such way that at infinity $\beta(r)^2 {\longrightarrow}(\beta(r \rightarrow {\infty}))^2= -b^2 $.
In the Minkowski space $(1+2)D$ we can observe that,
\begin{equation}
\label{LAR} \beta(r)^2=\beta_0(r)^2-\beta_r(r)^2-\beta_{\theta}(r)^2 \hspace{1cm} \mbox{and so} \hspace{1cm} \lim_{r \rightarrow \infty} \beta(r)^2= b_0^2-b_r^2-b_{\theta}^2 = - b^2 \, .
\end{equation}
Note that the magnitude of vector $\beta^{\mu}(r)$ must be a space-like vector.
As the gauge field $A^{\mu}$ is a function that exclusively depends on the variable $r$, and the assumption (\ref{AeB}) implies to $\epsilon^{\mu\nu\kappa}(B_{\nu}^*B_{\kappa}-B_{\nu}B_{\kappa}^*)=0$,
the equation of motion (\ref{em})  assumes, in the static configuration, the form:
\bn
\label{em1}
\partial_if^{\mu i}&=&  ie[B_{\nu}^{*} B^{\mu \nu} - B_{\nu}(B^{\mu \nu})^{*} + (B^{\mu})^*(\partial_{\nu}B^{\nu}) - B^{\mu}(\partial_{\nu}B^{\nu})^{*}]-2e^2A^{\mu}B_{\nu}^{*}B^{\nu} .
\en
The equation of motion (\ref{em0}) results in
\begin{equation}
\label{s}
D_{i}D^{i}B^{\mu}+2m\epsilon^{\mu\kappa\lambda }D_{\kappa}B_{\lambda} =\frac{\lambda}{2}\left(B_{\nu}^{*}B^{\nu}+b^2\right)B^{\mu}\, .
\end{equation}
Moreover the assumption (\ref{AeB}) indicates that the only non null component of $A^{\mu}$ vector is $A^{\theta}$, so the term $\partial_if^{\theta i}$ reduces to $\partial_rf^{\theta r} = \partial_r(\partial^{\theta} A^{r} - \partial^{r} A^{\theta}) =\partial_r(\nabla\times \vec{A})_k$.
The equation of motion  (\ref{em1}) for gauge field $A(r)$ can be written down as,
\begin{equation}
\label{Nielsen}
r^2\frac{d^2A(r)}{dr}+r\frac{dA(r)}{dr}+\left[2e^2\beta(r)^2 r^2-1\right]A(r) =
2en\beta(r)^2 r \, ,
\end{equation}
and the equation of motion  (\ref{s}) for Vector matter field yields three equations,
\begin{eqnarray}
\label{Nielsen1} \frac{1}{r}\frac{d}{dr}\left[r\frac{d\beta_{0}(r)}{dr}\right]-\left[\left(\frac{n}{r}-eA(r)\right)^2 - \frac{\lambda}{2}\left(\beta(r)^2+b^2\right) \right]\beta_{0}(r)&=&2m\left[\frac{d\beta_{\theta}(r)}{dr}-\left(\frac{in}{r}-ieA(r)\right)\beta_r(r)\right] \,, \\
\label{Nielsen2} \frac{1}{r}\frac{d}{dr}\left[r\frac{d\beta_{r}(r)}{dr}\right]-\left[\left(\frac{n}{r}-eA(r)\right)^2 - \frac{\lambda}{2}\left(\beta(r)^2+b^2\right)\right]\beta_{r}(r)&=&-2m\left(\frac{in}{r}-ieA(r)\right)\beta_0(r) \, , \\
\label{Nielsen3} \frac{1}{r}\frac{d}{dr}\left[r\frac{d\beta_{\theta}(r)}{dr}\right]-\left[\left(\frac{n}{r}-eA(r)\right)^2 - \frac{\lambda}{2}\left(\beta(r)^2+b^2\right) \right]\beta_{\theta}(r)&=&2m\frac{d\beta_0(r)}{dr} \, .
\end{eqnarray}
We might now establish the solutions for the following vortex line. Recall that the vortex lines occur for the spontaneous LV of space-like matter vector, for $b^2<0$. We write $\beta(r)^2 = -|b|^2$ and plug it in the equation (\ref{Nielsen}) that follows,
\begin{equation}
 \label{Nielsen_B}
r^2\frac{d^2A(r)}{dr}+r\frac{dA(r)}{dr}-\left[2e^2|b|^2 r^2+1\right]A(r) = - 2en|b|^2 r \, .
\end{equation}
In this case, where the vector field is space-like vector, the electromagnetic potential vector and magnetic field are similar to the Nielsen-Olesen scalar case \cite{ryder,Nielsen}, where we have,
\begin{eqnarray}
A(r)&=& \frac{n}{er}+
\frac{C}{e}K_{1}\big(\sqrt{2}|eb|r)
\hspace*{0.4cm} \overrightarrow{ r \rightarrow \infty } \hspace*{0.4cm}
\frac{n}{er}
+\frac{C}{e}\sqrt{\frac{\pi}{2\sqrt{2}|eb|r}} \,\,e^{-\sqrt{2}|eb|r}  \, ,\\
\label{magnetic field_2}
H(r)&=& C\sqrt{2}\,|b|\,K_{0}\big(\sqrt{2}|eb|r)
\hspace*{0.4cm} \overrightarrow{ r \rightarrow \infty } \hspace*{0.4cm}
C\sqrt{2}\,|b|\,\sqrt{\frac{\pi}{2\sqrt{2}|eb|r}}\,\,e^{-\sqrt{2}|eb|r}\,,
\end{eqnarray}
where $K_0$ and $K_1$ are modified Bessel functions and we have observed that the magnetic field decays rapidly for large values of $r$.
This solutions corresponding to the magnetic flux in a core of vortex lines that we can obtain an approximate mathematical shape with equations (\ref{Nielsen1}), (\ref{Nielsen2}) and (\ref{Nielsen3})
It can be done by eliminating the factors
$\left[\left(\frac{n}{r}-eA(r)\right)^2 - \frac{\lambda}{2}\left(\beta(r)^2+b^2\right)\right]$
and $\left(\frac{in}{r}+ieA(r)\right)$ (see Appendix A). So we obtain that
\begin{equation}
\frac{1}{r}\frac{d}{dr}\left[r\frac{d\beta_{0}(r)}{dr}\right]-\frac{\beta_{r}}{r\beta_{0}}\frac{d}{dr}\left[r\frac{d\beta_{r}(r)}{dr}\right]
+\chi\left(\frac{\beta_{r}^2}{\beta_{0}}-\beta_{0}\right)=2m\frac{d\beta_{\theta}(r)}{dr}\,,
\label{betas}
\end{equation}
with
\begin{equation}
\label{betas2}
\chi = \frac{1}{r\beta_{\theta}}\frac{d}{dr}\left[r\frac{d\beta_{\theta}(r)}{dr}\right] - \frac{2m}{\beta_{\theta}}\frac{d\beta_{0}(r)}{dr} \,.
\end{equation}
Note that Eqs. (\ref{betas}) and (\ref{betas2}) are general expressions that connect the components of the vector $\beta_{\mu}(r)$. We can now look for a solution
among some possible ones described in the Appendix A. Thus we can analyze the following vector:
in the form
\begin{equation}
\label{betas3}
\beta_{\mu}(r)=(\beta_{0}(r),\beta_{r}(r),\beta_{\theta}(r))=(b_{0}f(r),b_{r},b_{\theta}f(r))\,,
\end{equation}
where $\lim_{r\rightarrow\infty}f(r)=1$ and $\lim_{r\rightarrow 0}f(r)=0$. Inserting (\ref{betas3}) in Eq. (\ref{betas}) it results in
\begin{equation}
\label{equation beta 1}
\frac{d^2f(r)}{dr^2} + \left(\frac{1}{r}+2m\Lambda\right)\frac{df(r)}{dr} = 0,
\end{equation}
where $m$ is the topological mass and $\Lambda$ is equal to $\Lambda = \dfrac{b_{0}}{b^{\theta}}$.
Therefore, we can establish the general solution for Eq. \eqref{equation beta 1} as
\begin{equation}
\label{beta4}
f(r)=c_{1}+c_{2}Ei(- m\Lambda r)\,,
\end{equation}
where $c_{1}$ and $c_{2}$ are constants to be determined and $Ei(x)$ is the exponential integral function. Assuming $\lim_{r\rightarrow\infty}f(r)=1$, $\lim_{r\rightarrow 0}f(r)=0$, we can go to a step further, considering regions where $r\gg 0$, which leads \eqref{beta4} to the equation
\begin{equation}
\label{beta5}
f(r)=1-e^{-m\Lambda r}.
\end{equation}
We can obtain the square vector norm
\begin{equation}
\label{betas5}
\beta(r)^2=\beta_{\mu}\beta^{\mu}=(b^{0})^2 (1-e^{-m\Lambda r})^2 - (b^{r})^2 - (b^{\theta})^2 (1-e^{-m\Lambda r})^2 \,,
\end{equation}
that evidently tends to the value $\beta(r)^2=\beta_{\mu}\beta^{\mu}=(b^{0})^2 - (b^{r})^2 - (b^{\theta})^2 $ when $r\rightarrow \infty$.
The result (\ref{beta5}) depends on the topological mass $m$ as well as the planar space-like LV parameter $ -|b^2|$. We can observe that the vector matter field $B_{\mu}=\beta_{\mu}(r)e^{in\theta}$, which has a U(1) symmetry group, clearly has a boundary value in 2-dimensional space. This boundary is certainly a vortex circle $S^{1}$ for which $\theta$ ranges from $0$ to $2\pi$.

%%%%%%%%%%%%%%%%%%%%%%%%%%%%%%%%%%%%
%%%%%%%%%%%%%%%%%%%%%%%%%%%%%%%%%%%%
%%%%%%%%%%%%%%%%%%%%%%%%%%%%%%%%%%%%
%%%%%%%%%%%%%%%%%%%%%%%%%%%%%%%%%%%%
%%%%%%%%%%%%%%%%%%%%%%%%%%%%%%%%%%%%
%%%%%%%%%%%%%%%%%%%%%%%%%%%%%%%%%%%%
%%%%%%%%%%%%%%%%%%%%%%%%%%%%%%%%%%%%
%%%%%%%%%%%%%%%%%%%%%%%%%%%%%%%%%%%%
%%%%%%%%%%%%%%%%%%%%%%%%%%%%%%%%%%%%
%%%%%%%%%%%%%%%%%%%%%%%%%%%%%%%%%%%%
%%%%%%%%%%%%%%%%%%%%%%%%%%%%%%%%%%%%
%%%%%%%%%%%%%%%%%%%%%%%%%%%%%%%%%%%%

\section{Conclusion}

In this article, we present a model in which topological defects emerge from a vector matter field in $(1+2)D$, extending the standard concept of a scalar soliton. To this goal, we consider the vector field derived from the string model, which represents the LV effect, as observed by Kostelecky and Samuel \cite{Kostelecky_Samuel}. Based on these works and from the Maxwell-Chern-Simons \cite{Dunne} and topological defects \cite{Seifert} we construct a model to determine the topological features of the vector model in $(1+2)D$, as well as the vacuum LV parameter.

Even when introducing a Chern-Simons topological charge (mass) term as a proposal to stabilize the energy, we arrive at the same conclusion that Seifert \cite{Seifert}: a vector field domain wall does not satisfy the necessary energy condition, rendering its existence unlikely for a vector field of matter due to its negative surface energy density. In our analysis of the domain wall in $(1+2)D$ spacetime we find  that such solutions are unstable and thus physically inadmissible.
For the case of line vortex, arising from the spontaneous LV of a spacelike matter vector field, it becomes possible to stabilize the energy through the interaction with a gauge field, as described by the equations (\ref{DC}) and (\ref{PG}). Additionally, we obtain a matter vector field solution that yields  finite positive energy in the static regime.
The vortex lines solution in $(1+2)D$ spacetime proves well-suited to this dimensionality, analogous to the scalar Nielsen-Olesen vortex lines \cite{ryder, Vilenkin}, which describe a magnetic vortex decaying from the string's center.

%%%%%%%%%%%%%%%%%%%%%%%%%%%%%%%%%%%%%%%%%%%%%
%%%%%%%%%%%%%%%%%%%%%%%%%%%%%%%%%%%%%%%%%%%%%
%%%%%%%%%%%%%%%%%%%%%%%%%%%%%%%%%%%%%%%%%%%%%
%%%%%%%%%%%%%%%%%%%%%%%%%%%%%%%%%%%%%%%%%%%%%

The coupling of LV vector fields to gauge potentials is a pivotal area of research, though the underlying motivations diverge significantly based on the theoretical framework. This work examines this distinction by contrasting two primary approaches. The first, as explored in the context of gravitational model-building \cite{Lehum:2024}, utilizes a metric-affine formulation to demonstrate how non-metricity sourced by a bumblebee field induces novel couplings to an electromagnetic field, with the principal objective of computing one-loop quantum corrections and deriving the resulting low-energy effective action. The second approach, which we develop here concerning topological defects, operates within a lower-dimensional setting. In this paradigm, the gauge coupling is essential not for radiative effects but for stabilizing the energy of classical topological solutions such as vortices and domain walls ensuring magnetic flux quantization, and analyzing the influence of a topological Chern-Simons mass. This juxtaposition underscores a key insight: the mechanism of spontaneous LV serves as a versatile foundation for addressing disparate physical questions, spanning from quantum gravity phenomenology to the structure and stability of topological defects in field theory.
%%%%%%%%%%%%%%%%%%%%%%%%%%%%%%%%%%%%%%%%%%%%%
%%%%%%%%%%%%%%%%%%%%%%%%%%%%%%%%%%%%%%%%%%%%%
%%%%%%%%%%%%%%%%%%%%%%%%%%%%%%%%%%%%%%%%%%%%%
Further investigations could explore the role of more complex topological defect configurations, such as hybrid defects or higher-dimensional extensions, in LV frameworks. Additionally, numerical simulations could provide deeper insights into the stability and dynamics of these structures under varying symmetry-breaking conditions. The interplay between quantum effects and topological defects in low-dimensional systems also remains an open avenue for research, particularly in condensed matter analogs of these field-theoretic models.

%\section*{CRediT authorship contribution statement }

%\textbf{Luiz P. Colatto}: Writing --review \& editing, Writing --original draft, Visualization, Validation, Methodology, Investigation, Formal analysis, Conceptualization. \textbf{Andre L.A. Penna}: Writing --review \& editing, Writing --original draft, Visualization, Validation, Methodology, Investigation, Formal analysis, Conceptualization. \textbf{Wytler C. dos Santos}: Writing --review \& editing, Writing --original draft, Visualization, Validation, Methodology, Investigation, Formal analysis, Conceptualization.

%\section*{Declaration of competing interest}

%The authors declare that they have no known competing financial interests or personal relationships that could have appeared to influence the work reported in this paper.

\section{Acknowledgements}

\noindent
The authors would like to thank to Prof. J.A. Helay\"el-Neto for discussions and the kind hospitality at the CBPF.

\appendix

\section{Solutions to space-like vector field in $(1+2)D$ spacetime}

We are now looking for some approximate solution dealing with topological mass for the Bumblebee fields components $\beta_{0}(r)$,\,\,$\beta_{r}(r)$,\,\,$\beta_{\theta}(r)$. To this goal it requires to focus on solutions in eqs. \eqref{Nielsen1}, \eqref{Nielsen2} and \eqref{Nielsen3}
with topological mass at $r<\infty$.
\bn
\label{Nielsen4} \frac{1}{r}\frac{d}{dr}\left[r\frac{d\beta_{0}(r)}{dr}\right]-\left[\left(\frac{n}{r}-eA(r)\right)^2 - \frac{\lambda}{2}\left(\beta(r)^2+b^2\right)\right]\beta_{0}(r)=2m\left[\frac{d\beta_{\theta}(r)}{dr}-\left(\frac{in}{r}-ieA(r)\right)\beta_r(r)\right] \, ,
\en
\bn
\label{Nielsen5} \frac{1}{r}\frac{d}{dr}\left[r\frac{d\beta_{r}(r)}{dr}\right]-\left[\left(\frac{n}{r}-eA(r)\right)^2 - \frac{\lambda}{2}\left(\beta(r)^2+b^2\right) \right]\beta_{r}(r)=-2m\left(\frac{in}{r}-ieA(r)\right)\beta_0(r) \, ,
\en
\bn
\label{Nielsen6} \frac{1}{r}\frac{d}{dr}\left[r\frac{d\beta_{\theta}(r)}{dr}\right]-\left[\left(\frac{n}{r}-eA(r)\right)^2 - \frac{\lambda}{2}\left(\beta(r)^2+b^2\right) \right]\beta_{\theta}(r)=2m\frac{d\beta_0(r)}{dr} \, .
\en
In the equation (\ref{Nielsen6}) we isolate the term  $\left[\left(\frac{n}{r}-eA(r)\right)^2 - \frac{\lambda}{2}\left(\beta(r)^2+b^2\right) \right]$, where we obtain,
\bn
\label{Nielsen7} \left[\left(\frac{n}{r}-eA(r)\right)^2 - \frac{\lambda}{2}\left(\beta(r)^2+b^2\right) \right] = \frac{1}{\beta_{\theta}(r)}\frac{1}{r}\frac{d}{dr}\left[r\frac{d\beta_{\theta}(r)}{dr}\right]- \frac{1}{\beta_{\theta}(r)}2m\frac{d\beta_0(r)}{dr} = \chi\, .
\en
Then multiply the equation (\ref{Nielsen4}) by $\beta_{0}(r)$ and multiply the equation (\ref{Nielsen5}) by $\beta_{r}(r)$ and then subtract them. Therefore we obtain:
\begin{equation}
 \beta_{0}\frac{1}{r}\frac{d}{dr}\left[r\frac{d\beta_{0}(r)}{dr}\right]-\chi(\beta_{0})^2 -\beta_{r}\frac{1}{r}\frac{d}{dr}\left[r\frac{d\beta_{r}(r)}{dr}\right] +\chi (\beta_{r})^2 = 2m \beta_{0}\left[\frac{d\beta_{\theta}(r)}{dr}\right]
\end{equation}
Now dividing the above equation by $\beta_{0}$ we obtain:
\begin{equation}
\label{equation_beta_01}
\frac{1}{r}\frac{d}{dr}\left[r\frac{d\beta_{0}(r)}{dr}\right]-\frac{\beta_{r}}{r\beta_{0}}\frac{d}{dr}\left[r\frac{d\beta_{r}(r)}{dr}\right]
+\chi\left(\frac{\beta_{r}^2}{\beta_{0}}-\beta_{0}\right)=2m\frac{d\beta_{\theta}(r)}{dr}\,,
\end{equation}
with
\begin{equation}
\label{equation_beta_02}
\chi = \frac{1}{r\beta_{\theta}}\frac{d}{dr}\left[r\frac{d\beta_{\theta}(r)}{dr}\right] - \frac{2m}{\beta_{\theta}}\frac{d\beta_{0}(r)}{dr} \,.
\end{equation}

These equations can be solved when we have $\beta_{\mu}(r) = \beta_{\mu}(f(r))$, with the following conditions that obey vacuum conditions,
\begin{equation}
\label{condition_vacuum}
 \lim_{r\rightarrow\infty} \beta_{\mu}(f(r)) = b_{\mu}, \hspace*{1cm} \mbox{or} \hspace*{1cm} \lim_{r\rightarrow\infty} |f(r)| = 1.
\end{equation}
Let us expose seven simple possible conditions for the vector $\beta_{\mu}(f(r))$ as:
\begin{enumerate}

 \item $\beta_{\mu}(r) = (b_{0}f(r),b_{r},b_{\theta})$;
  \item $\beta_{\mu}(r) = (b_{0},b_{r}f(r),b_{\theta})$;
  \item $\beta_{\mu}(r) = (b_{0},b_{r},b_{\theta}f(r))$;
  \item $\beta_{\mu}(r) = (b_{0}f(r),b_{r}f(r),b_{\theta})$;
  \item $\beta_{\mu}(r) = (b_{0}f(r),b_{r},b_{\theta}f(r))$;
  \item $\beta_{\mu}(r) = (b_{0},b_{r}f(r),b_{\theta}f(r))$;
  \item $\beta_{\mu}(r) = (b_{0}f(r),b_{r}f(r),b_{\theta}f(r))$.
 \end{enumerate}
and as previously stated, the $b_{0}$, $b_{r}$ and $b_{\theta}$ are vacuum expectation constant values. After using each of the possible proposals for the vector field $\beta_{\mu}(r)$ in the equations (\ref{equation_beta_01}) and (\ref{equation_beta_02}),
we get that the solution which is stable and suitable is proposed number 5.
For $\beta_{\mu}(r) =  (b_{0}f(r),b_{r},b_{\theta}f(r))$ the differential equation (\ref{equation_beta_01}) results in
\begin{equation}
 \frac{b_{0}}{r} \frac{d}{dr}\left(r \frac{df(r)}{dr} \right) + \left[ \frac{1}{f(r) r} \frac{d}{dr}\left(r \frac{df(r)}{dr}\right) -2m \frac{b_{0}}{f(r)b_{\theta}} \frac{df(r)}{dr}\right]\left(\frac{b_{r}^2}{b_{0} f(r)} - b_{0}f(r) \right) = 2m~b_{\theta} \frac{df(r)}{dr}. \nonumber
\end{equation}
Simplifying the equation above we then obtain:
\begin{equation}
\label{equation_beta_06.1}
 \frac{b_{r}^2}{b_{0} (f(r))^2} \left(\frac{1}{r} f'(r) + f''(r) - 2m \frac{b_{0}}{b_{\theta}}f'(r) \right) + 2m \frac{b_{0}^2}{b_{\theta}} f'(r) =  2m~b_{\theta}~f'(r).
\end{equation}
If by chance we choose $b_{r}=0$ the above equation reduces to:
 \begin{equation}
  2m~\frac{b_{0}^2}{b_{\theta}} f'(r) = 2m~b_{\theta}f'(r), \nonumber
 \end{equation}
which only results in $|b_{0}|= |b_{\theta}|$ or $b_{0}= \pm b_{\theta}$ for any arbitrary function $f(r)$.

Now we can assume $b_{r}$ a non-zero constant with $b_{0}^2= b_{\theta}^2$ to reduce the differential equation (\ref{equation_beta_06.1}) as follows:
\begin{equation}
 \frac{d^2 f(r)}{dr^2} + \left(\frac{1}{r} - 2m \frac{b_{0}}{b_{\theta}}\right)\frac{df(r)}{dr} = 0. \nonumber
\end{equation}
We must have $b^{\theta}$ definite positive. The vector $B^{\mu} = \beta^{\mu}(r) e^{in\theta}$ has $\beta^{\mu}(r)$ defined as module. The spacial covariant components are $b_{r} = - b^{r}$ and  $b_{\theta} = - b^{\theta}$. So, the above differential equation with $\Lambda = \dfrac{b_{0}}{b^{\theta}}$ can be write as
\begin{equation}
\label{equation_beta_06.2}
 \frac{d^2 f(r)}{dr^2}  + \left(\frac{1}{r} + 2m\Lambda \right)\frac{df(r)}{dr}=0\,,
\end{equation}
where $b_r$ is a non-zero constant. Therefore, we can establish the general solution for above differential equation as
\begin{equation}
\label{equation_beta_06.3}
f(r)=c_{1}+c_{2}Ei(-2m\Lambda r)\,,
\end{equation}
where $c_{1}$ and $c_{2}$ are constants to be determined and $Ei(x)$ is the exponential integral function. Assuming $\lim_{r\rightarrow\infty}f(r)=1$, $\lim_{r\rightarrow 0}f(r)=0$, we can go to a step further, considering regions where $r\gg 0$, which leads \eqref{equation_beta_06.3} to the equation
\begin{equation}
\label{equation_beta_06.4}
f(r)=1-e^{-m\Lambda r}\,,
\end{equation}
such that in order to have a stable and convergent solution, $\Lambda =1$, thus we have ${b_{0}} = {b^{\theta}} =  - {b_{\theta}} $.
For  matter space-like vector field the solution (\ref{equation_beta_06.4}) seems to be the most suitable.

%\section*{Data availability}

%No data was used for the research described in the article.

%\section{References}

%\nocite{*}

%\bibliographystyle{elsart-num}
%\bibliographystyle{unsrt}
%\bibliographystyle{unsrtnat}
\bibliographystyle{elsarticle-num-names}
%\biboptions{numbers}

\bibliography{Biblio.bib}

\end{document}